%% file: main.tex
\documentclass[twoside, 11pt]{article}
\usepackage[left=1in,top=1in,right=1in,bottom=1.5in]{geometry} 
\usepackage[reqno]{amsmath}
\usepackage[T1]{fontenc}
\usepackage{amssymb,amsmath,graphicx,verbatim,url,verbatim,color,rotating,setspace,skak,multirow}

\usepackage[lofdepth,lotdepth]{subfig}

\usepackage{enumitem}
\setenumerate{noitemsep}
\usepackage{tikz}
\usetikzlibrary{shapes,arrows,calc}
\usetikzlibrary{decorations.pathreplacing}
\usepackage{todonotes}

\usepackage{color}\definecolor{spot}{rgb}{0.6,0,0}

\usepackage[pdftex, bookmarksopen=true, bookmarksnumbered=true,
  pdfstartview=FitH, breaklinks=true, urlbordercolor={0 1 0},
  citebordercolor={0 0 1}, colorlinks=true,
            citecolor=spot,
            linkcolor=spot,
            urlcolor=spot,
            pdfauthor={Authors},
pdftitle={Title}]{hyperref}

\title{\textsf{The Privacy-preserving Padding Problem: Non-negative Mechanisms for Conservative Answers with Differential Privacy}}

\author{Benjamin M.\ Case\thanks{Research Scientist, Facebook; \texttt{bmcase@fb.com}}, James Honaker\thanks{Research Scientist, Facebook; \texttt{jameshonaker@fb.com}}, Mahnush Movahedi\thanks{Research Scientist, Facebook; \texttt{mahnush@fb.com}}  }

\begin{document}

\maketitle

\begin{abstract}
Differentially private noise mechanisms commonly use symmetric noise distributions.  This is attractive both for achieving the differential privacy definition, and for unbiased expectations in the noised answers.  However, there are contexts in which a noisy answer only has utility if it is conservative, that is, has known-signed error, which we call a padded answer.  Seemingly, it is paradoxical to satisfy the DP definition with one-sided error, but we show how it is possible to bury the paradox into approximate DP's  $\delta$ parameter.  We develop a few mechanisms for one-sided padding mechanisms that always give conservative answers, but still achieve approximate differential privacy. We show how these mechanisms can be applied in a few select areas including making the cardinalities of set intersections and unions revealed in Private Set Intersection protocols differential private and enabling multiparty computation protocols to compute on sparse data which has its exact sizes made differential private rather than performing a fully oblivious more expensive computation. 
\end{abstract}

\input{intro}
\input{oneside}

\section{Mechanisms}\label{s:mech}
\input{mechanismLaplace}

\input{mechanismGeometric}

\input{mechanismPoisson}

\input{mechanismNegBin}
\input{mechanismHeuristics}

\input{applicationPSI}

\input{applicationDPHist}

\input{priorwork}
\input{conclusion}

\newpage
\appendix
\input{appendix}

\clearpage

\bibliographystyle{IEEEtran}
\bibliography{positive}

\end{document}

%% file: intro.tex
\section{Motivation}

\begin{quote}
\textbf{Scenario:} Jane wants to kill her husband Lord Edgware.  She has an elaborate false alibi but is uncertain if Lord Edgware will be attending a particular dinner party.  The butler is to be informed how many people to provision for; Jane may have already learned which other people are attending.  Protective of his guests, the host sends the butler a differentially private count, so that if Jane intercepts the message, Jane can not learn if Lord Edgware will be attending so as to murder him.  However, the butler does as is told, and it would be an impropriety for there not to be enough provisions for everyone in attendance.  It is therefore essential that the number given to the butler is both differentially private (so no one is murdered), \emph{and never be less than the true number of attendees} (so every attendee can be accomodated).  
\end{quote}

The solution to this problem has direct applications to a number of scenarios where a differentially private answer is needed to preserve privacy, but the answer must be \emph{conservative} and thus any noise added must be non-negative and never give an answer less than the true value.  Conservative answers are often desirable in contexts where the utility loss from error is more heavily weighted on one side.  
One might carry a conservative amount of cash for a purchase, because having left over cash has no cost, but having insufficient cash means the purchase is not completed.  


There are a number of applied contexts where a non-negative DP noise mechanism would provide an important solution and basic building block:
\begin{itemize}
\item As a computational example, Kellaris et al.~\cite{kellaris2017accessing} consider the problem where an adversary is monitoring the size of traffic between a secure database and an owner subsetting observations by range queries.  Even if the queries transmitted and the rows returned are encrypted, if the adversary observes the true distribution of the size of returns, they are eventually able to reconstruct the distribution of the dataset.  Sending query responses with differentially private numbers of rows defeats the adversary.  However while sending a larger than correct subset (that is, padding with removable fake data) has low cost, sending a smaller than correct subset (by omitting observations that satisfy the query) has a high utility cost. 

\item Similarly, private set intersection (PSI) that enable computation on the intersection of two private sets are important building block of secure solutions \cite{PJC2019,buddhavarapu2020private,Pinkas2019circuitlinear} which allows two parties to locate the intersection of their data without directly revealing to either party which observations the other has in common.  However, many PSI protocols either reveal the cardinalities of the set intersections and unions as an intended output or as an additional leakage to the intended output; this opens up attacks such as differencing for membership inference.  Padding the set intersection and union with dummy records sampled according to non-negative DP machanisms can solve this problem by making the cardinalities differentially private from the view of both parties. 

\item In MPC games for conversion measurement, such as in \cite{movahedi2021privacy} a mechanism for positively padded DP histograms would provide a means to prevent the leakage of the number of conversions without computationally inefficient padding of all conversion event data to the upper bound of conversion count.  

\item More generally, side-channel attack through timing or data access patterns have been a long recognized problem in both MPC and differentially private systems \cite{haeberlen2011differential} when such side-channels are considered parts of the output that need to meet the differential privacy definition.  Methods suggested to avoid such attacks have involved making all computations \emph{constant-time} but using our mechanisms to pad compute or storage with a stocastically drawn strictly positive DP waiting time we show can be a much more efficient solution.

\item In the foundational question of releasing a differentially private mean on a set of unknown size, Covington et al.~\cite{covington2020unknown} show a setting where lowest mean squared error is achieved when you can privately undercount the set (which is an example of needing a conservative answer with non-positive error).  
\end{itemize}

In this work, we present a general approach to creating noise mechanisms with error with guaranteed sign, and solidify three such mechanisms.  We show how this provides a padding solution to PSI intersection and union leakages, as well as one example of overcoming distribution and differencing attacks on MPC implementations in the context of conversion measurement.

%% file: oneside.tex
\section{One-sided Noise}

We are going to consider privatizing noise addition mechanisms, where the noise $z$ is drawn from some distribution $p$ with no support on the negative numbers,  that is:
\begin{align}
M(X) &= f(X) \pm z; \qquad z \sim p: p(s)=0 \quad \forall \ s<0
\end{align}
Statistical estimators whose error is of a guaranteed sign are called \emph{conservative}, and can be desirable in certain applied contexts.  We use $\pm$ to highlight that we can guarantee the sign of the error in either direction; we can add $z$ or subtract $z$.  Throughout this paper, we write as if we are adding $z$ so as to guarantee non-negative errors and \emph{overvalued} answers, but everything holds if instead we need to subtract $z$ to guarantee non-positive errors in a differentially private \emph{undervalued} answer.

Let us first highlight the key dilemma of any such mechanism.  Imagine two neighbouring datasets $X$ and $X'$ which happen to have $f(X) < f(X')$.  We release a differentially private answer $M$.  Then whenever we observe a  release $M< f(X')$ we know we must be in dataset $X$ and never in $X'$; the mechanism $M$ can never give an answer below $f(X')$ if $X'$ is the private data.  This means we can leak with certainty which state of the world we are in, which in turn violates differential privacy.  For example, if we are doing a counting query, and know we have either dataset $X$ which has count 100, or neighboring dataset $X'$ which has count 101, and we use a non-negative noise distribution, then whenever we see an answer of 100, we know we were in dataset $X$.  To remove this probability we are going to have to move to approximate differential privacy and bury all $M< f(X')$ into $\delta$. We now consider distributions $p$ that satisfy $(\epsilon,\delta)$-differential privacy in this way.

%% file: mechanismLaplace.tex
\subsection{Truncated Laplace}

In the continuous case, consider a differentially private mechanism, $M$, that releases answers to a query, $f$, on a dataset, $X$, with truncated Laplace noise as:
\begin{align}
M(X) &= f(X) + z; \quad z\sim p_{TruncatedLaplace}(b,\mu)\\
Pr_{TruncatedLaplace}(x | b, \mu ) &= \left\{
\begin{array}{ll}
      0 & x < 0 \\
      \frac{A}{2b}\ \textrm{exp} \Big(-\frac{|x-\mu|}{b}\ \Big)
 & x \geq 0 \\
\end{array} 
\right. 
\quad \textrm{where  }b = \frac{\Delta}{\epsilon}
\end{align}
with mode $\mu$, and shape parameter $b$ (variance $2b^2$ when untruncated), and where all the mass below 0 has been truncated, so we need a normalizing constant $A$, to bring the total probability mass back to one.  We can solve for $A$ (assuming $x<\mu$) as:
\begin{align}
A &= \Big( 1- \int_{-\infty}^0 \frac{1}{2b}\ \textrm{exp} \Big(-\frac{|x-\mu|}{b}\ \Big)  dx \Big)^{-1} =  \Big( 1 - \frac{1}{2}\ \textrm{exp} \Big(\frac{-\mu}{b}\Big) \Big) ^{-1} \quad \ 
\end{align}
Note because $A$ works here by inflating the entire distribution by a constant factor, in any ratio, these factors cancel.  Thus the truncated Laplace continues to obey the differential privacy definition which is itself defined in terms of ratios.  (This is only true when we truncate the distribution at the same distance from the mean for all datasets.)

Whereas the Laplace satisfies pure $\epsilon$-differential privacy, the truncated version is ($\epsilon$,$\delta$)-differentially private, so long as:
\begin{align}
\delta &\geq \int_0^{\Delta} \frac{A}{2b}\ \textrm{exp} \Big(-\frac{|x-\mu|}{b}\ \Big)  dx 
=  \frac{A}{2b} \textrm{exp}\Big(\frac{-\mu}{b}\Big) \int_0^{\Delta} \textrm{exp}\Big(\frac{x}{b}\Big) dx  
= \frac{A}{2} \textrm{exp}\Big(\frac{-\mu}{b}\Big)  \Big[ \textrm{exp}\Big(\frac{\Delta}{b}\Big) - 1 \Big] \label{delta}
\end{align}

\begin{figure}[tbhp!]
  \begin{center}
    \includegraphics[width=0.9\textwidth, trim= 0 0.5in 0 1.5in]{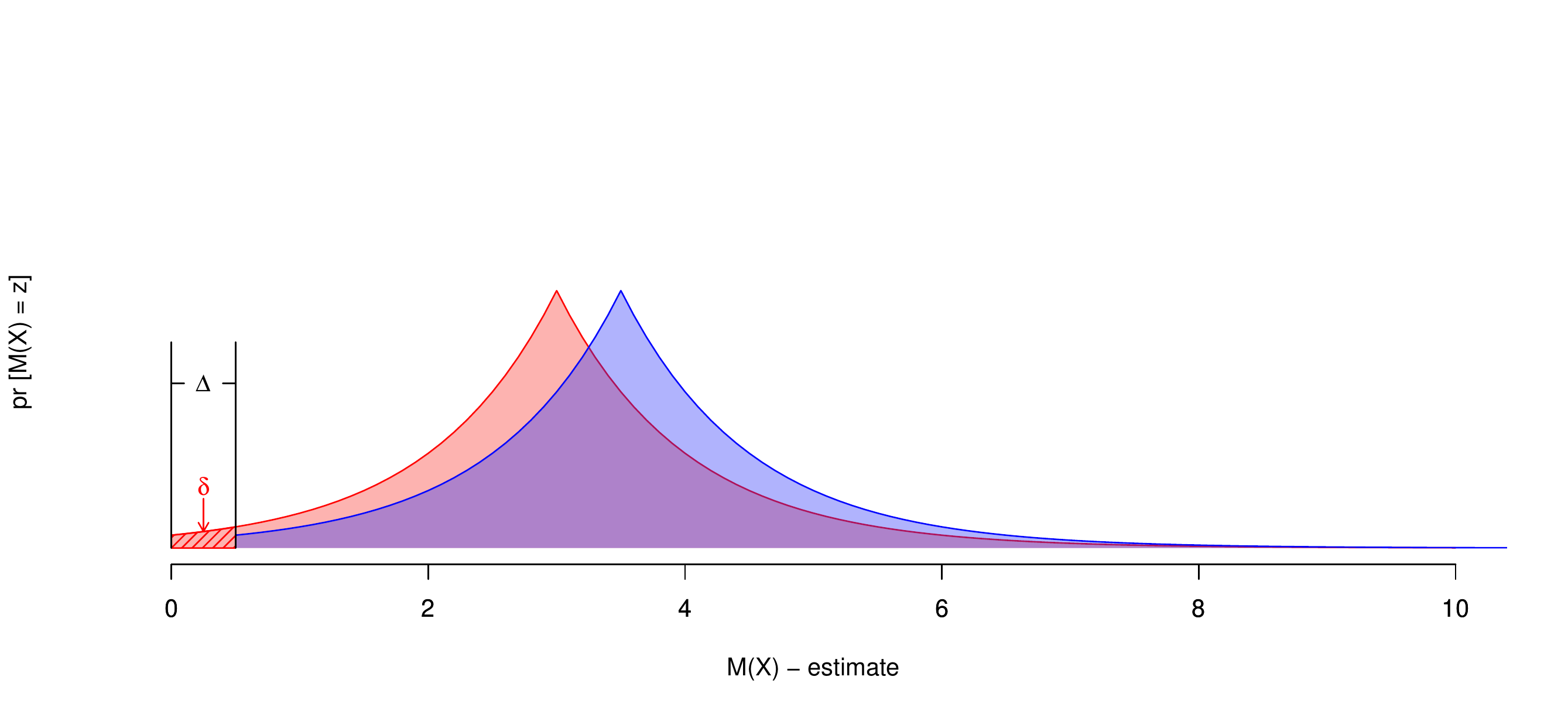}
    \end{center}
   \caption{\em Depicted are two truncated Laplace noise distributions with the same parameters, offset by some worst-case sensitivity $\Delta$.  The shaded region (exaggerated for visibility) shows the region with support in only one distribution.  Releases in this part of the noise distribution, can potentially reveal which of two neighbouring datasets are true.  That probability mass needs to be covered by $\delta$.}
 \label{f:laplace}
\end{figure}

\begin{figure}[tbhp!]
  \begin{center}
    \includegraphics[width=0.9\textwidth, trim= 0 0.5in 0 1.5in]{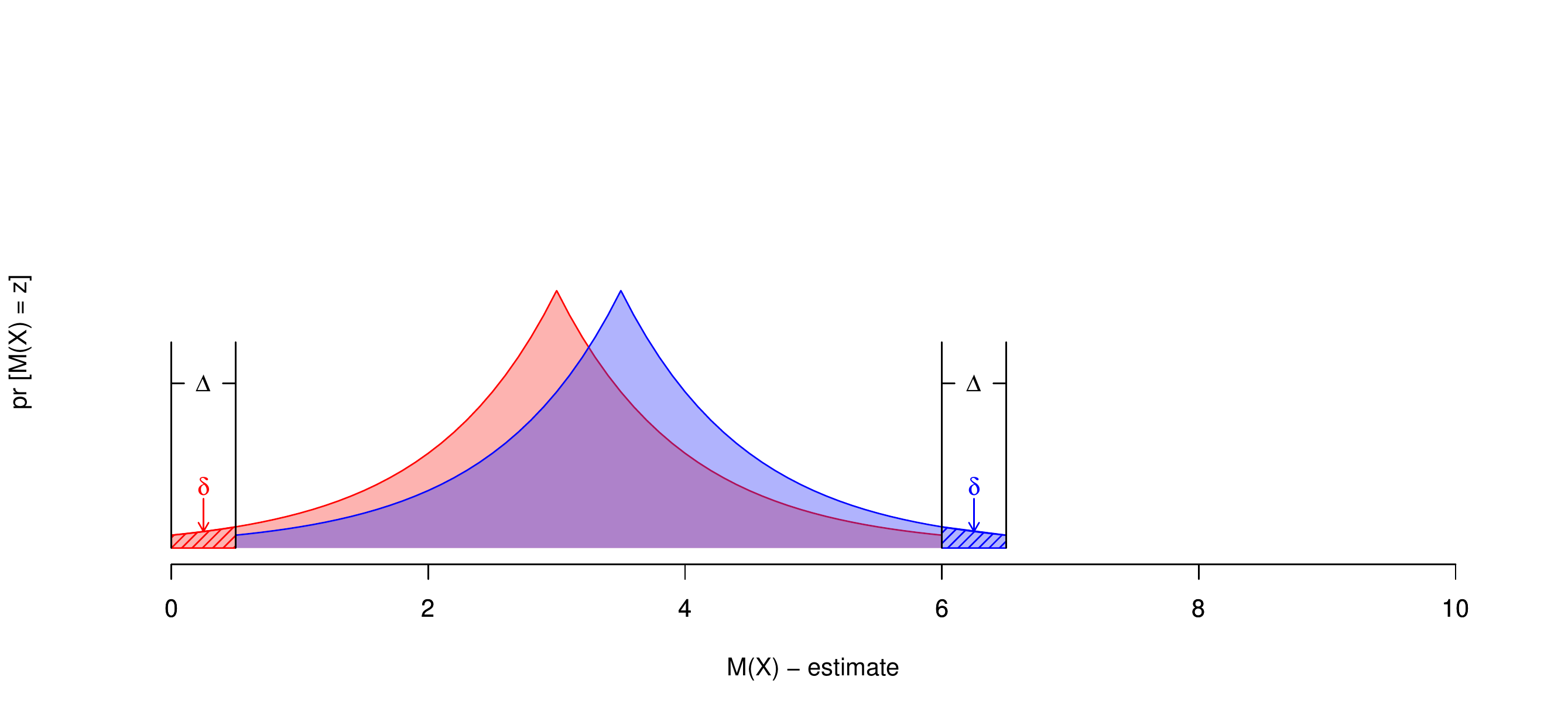}
    \end{center}
   \caption{\em When the Laplace is doubly and symmetrically truncated, two distributions that are offset by the worst-case sensitivity, $\Delta$, have two regions that have no overlap.  However, if there is a $p$ probability of being in the Red distribution, and a corresponding $1-p$ probability of being in the blue distribution, the total worst-case probability of a realized outcome in a non-overlapping region remains $\delta$.  This allows us to truncate both tails under approximate differential privacy, and retain a symmetric distribution.}
 \label{f:2trunc}
\end{figure}

A further note is that we can also symmetrically truncate the right tail of the distribution for no additional cost, as represented in Figure \ref{f:2trunc}.  While this means the regions where the distributions do not overlap apparently sum now to $2\delta$, we are either in dataset $X$, which fails with $\delta$ probability in the left tail, or in dataset $X'$ which fails with $\delta$ probability in the right tail.  In any state of the world, there remains only a $\delta$ chance of failure.  This slightly reduces the variance of the noise, and returns the distribution to symmetry which conveniently means the mode and mean are once again both at $\mu$.  This makes the revised inflationary constant:
\begin{align}
    A = \big( 1 - \textrm{exp} \Big(\frac{-\mu}{b}\Big) \big) ^{-1} \label{inflation}
\end{align}

We now require to know the mode as a function of $\epsilon$.  In conventional uses of the Laplace, $\mu=0$. In this use, we need $\mu$ to be sufficiently distant from zero that the integral for $\delta$ is satisfied.  At equality, Equation \ref{delta} together with \ref{inflation} gives us:

\begin{align}
\delta =   \frac{\textrm{exp}\Big(\frac{-\mu}{b}\Big)}{ 2 \big[ 1 - \textrm{exp} \Big(\frac{-\mu}{b}\Big) \big] }   \Big[ \textrm{exp}\Big(\frac{\Delta}{b}\Big) - 1 \Big]  &= 
\frac{\textrm{exp}\Big(\frac{-\mu\epsilon}{\Delta}\Big)}{ 2 - 2 \ \textrm{exp} \Big(\frac{-\mu\epsilon}{\Delta}\Big)  }   \Big( e^\epsilon - 1 \Big)
\Rightarrow \; \textrm{exp}\Big(\frac{-\mu\epsilon}{\Delta}\Big) 
= \frac{2\delta}{2\delta + e^\epsilon - 1} \nonumber \\ 
\Rightarrow \quad \mu &= 
 -\frac{\Delta}{\epsilon}\ \textrm{ln}\Big( \frac{2\delta}{2\delta + e^\epsilon - 1} \Big) 
\end{align}

Which ends up being quite succinct.  Note that $e^\epsilon\! - \!1 >0$ so the logarithmic term is always a negative number, giving us a strictly positive  $\mu$ as expected.

\subsubsection{Example}

If we have $\epsilon=0.5$ and $\delta=10^{-6}$ for a sum query with sensitivity $\Delta=1$ then this gives us $\mu=25.4$.  Thus we would expect to add 25.4 to the true sum.  Given the truncated symmetry of the noise distribution, we could add 0, and we would never add more than twice the mean, or 50.8.  At $\epsilon=1$ the expectation shrinks almost by half to  13.7.

%% file: mechanismGeometric.tex
\subsection{Truncated Geometric/Discrete Laplace}

Many of the use cases of padding involve adding integer counts of dummy users or events to mask true totals.  It is natural then to turn to distributions over the Whole numbers.  The geometric mechanism gives a discretized analog to the Laplace mechanism over the integers, and we can truncate this distribution similar to how we truncated the Laplace.  Symmetric about zero, Kotz et al.\ refer to this straightforwardly as the \emph{double geometric distribution} \cite[p.130]{Kotz2001}.   Consider a mechanism using this distribution:
\begin{align}
M(X) &= f(X) + z; \quad z \sim p_{DoubleGeometric}(n) \\
\textrm{Pr}_{DoubleGeometric}(x|n) &= Ae^{-\epsilon|n-x|}; \qquad x\in \{0,\ldots,2n\}
\end{align}
For some normalizing constant $0<A<1$ and some $n \in \mathbb{N}$.

\begin{figure}[tbhp!]
  \centering
	\subfloat[Example of the double geometric for $n=10$.]{\includegraphics[width=0.49\textwidth, trim= 0 0.2in 0 1.7in]{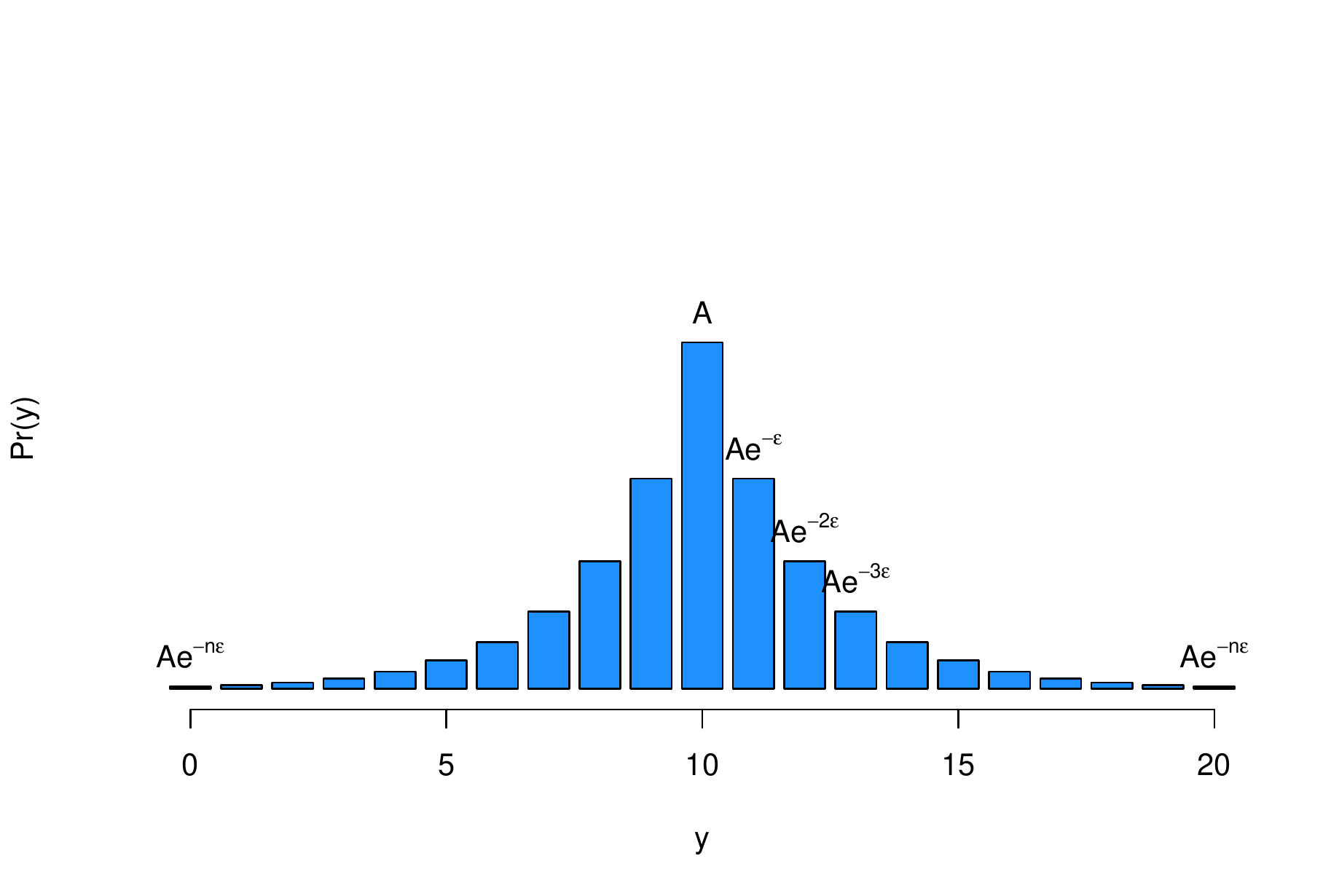}}
	\subfloat[Two geometric distributions offset by worst case sensitivity, $\Delta$, set here in this example to 1. The parameter $\delta$ has to cover either of the tail extremes that are not non-overlapping in distribution.]{\includegraphics[width=0.49\textwidth, trim= 0 0.2in 0 1.7in]{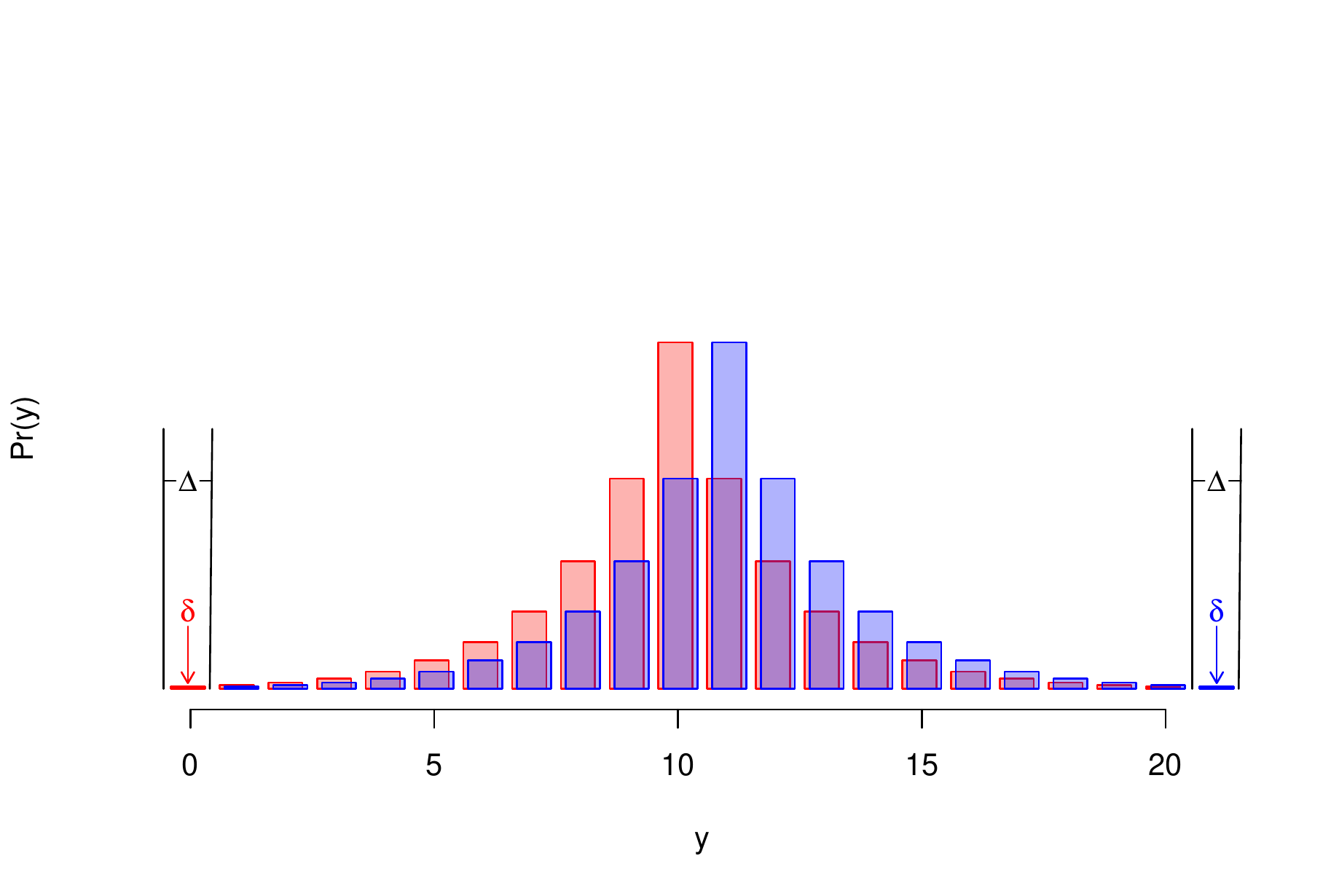}}
	\caption{\em Normalization and truncation of the geometric distribution.}
\end{figure}

As a probability this must sum to 1, which lets us solve for $A$.  Let $r=e^{-\epsilon}$.  Then we can rewrite as a classic geometric sequence as:
\begin{align}
1=\Big(2 A\sum_{k=0}^n e^{-k\epsilon}\Big) -A &= A\Big( -1 + 2\sum_{k=0}^n r^k \Big) 
= A\Big( -1 + 2 \frac{1 - r^{n+1} }{1-r} \Big) 
= A\Big( \frac{1 + r - 2r^{n+1} }{1-r} \Big) \nonumber \\
\Rightarrow \quad A &= \frac{1-r}{1 + r - 2r^{n+1} } = \frac{1-e^{-\epsilon}}{1 + e^{-\epsilon} - 2e^{-\epsilon(n+1)}} 
\end{align}

If we have sensitivity $\Delta \in \mathbb{Z}^+$, then we need $\delta$ to cover the tail as:
\begin{align}
\delta &\geq A \sum_{k=n-\Delta+1}^n  e^{-k \epsilon} 
\end{align}

For the common case of $\Delta=1$, such as in counting queries of users, at equality this simplifies (see appendix \ref{a:work}) to:
\begin{align} 
\delta = Ae^{-n \epsilon} = Ar^n &= \frac{r^n(1-r)}{1 + r -2r^{n+1}}  \\ 
\Rightarrow  \quad n &= \Big\lceil -\frac{1}{\epsilon}\ ln \Big(\frac{\delta (1 + r)}{1-r + 2r\delta}\Big) \Big\rceil \label{eq:expectN}
\end{align}

The geometric distribution is unwieldy analytically beyond $\Delta=1$ so it is natural to consider alternative distributions that do not have to be truncated to the non-negative integers, but naturally only have support there.  

\subsection{Example}

Consider we again target privacy-loss parameters of $\epsilon=0.5$ and $\delta=10^{-6}$ for a sum or counting query with sensitivity $\Delta=1$.  Then this would solve to $n=25$ as the expected value of the added noise, with the upper bound being 50.

%% file: mechanismPoisson.tex
\subsection{Poisson}

The Poisson distribution is the most common introductory statistical model of counting processes, and a useful starting point when we consider conservatively privatizing functions such as counting queries that have support over non-negative integers.  

Consider a mechanism with Poisson noise:
\begin{align}
M(X) &= f(X) + z; \quad z \sim p_{\textrm{Poisson}}(\lambda) \\
\textrm{Pr}_{Poisson}(y|\lambda) &= \frac{\lambda^y e^{-\lambda}}{y!}; \quad \lambda > 0.
\end{align}
for some constant rate parameter $\lambda$.  We assume $f(.) \in \mathbb{Z}$ has sensitivity $\Delta \in \mathbb{N}$.  Assume two neighbouring datasets are ordered for convenience such that $f(X) \leq f(X')$.  


In the region $k \geq f(X) + \Delta$ we know any two neighbouring datasets must have overlapping release distributions, whose ratio we can define as: 

\begin{align}
e^\epsilon \geq \max_{X,X'} \frac{M(X')}{M(X)} = \max_y \frac{\frac{\lambda^y e^{-\lambda}}{y!}}{\frac{\lambda^{y+\Delta} e^{-\lambda}}{(y+\Delta)!}} 
\quad \Rightarrow \quad \epsilon = -\Delta\textrm{log}(\lambda) + \max_y \prod_{i=1}^\Delta (y+i)
\end{align}
However, this rightmost term has no limit as $y\Rightarrow \infty$, so the right tail behavior of Poisson ratios does not collapse in a fashion that can converge to a limit.  One solution is that we can truncate the right tail of the Poisson so this limit isn't reached.  Another solution would be to determine the additional $\delta$ that is required to cover this tail behaviour, as is done for the Gaussian mechanism.  However, we instead shift now to distributions with improved tail behaviour.

%% file: mechanismNegBin.tex
\subsection{Negative Binomial}

While the Poisson has an intuitive generative form, the constraint that both the mean and the variance are directly determined by the same underlying parameter $\lambda$ can lead to inflexibility in applied settings.  There are, therefore, many practical generalizations of the Poisson that allow the mean and variance to be decoupled to independent parameters.  Moreover, the Poisson could not meet the differential privacy definition because it does not have a sub-exponential tail.  We turn now to a heavier tailed distribution on counts, the Negative Binomial, and show it can meet the approximate differential privacy definition as a noise mechanism.

Consider a mechanism with Negative Binomial noise as:
\begin{align}
M(X) &= f(X) + z; \quad z \sim p_{\textrm{NegBin}}(\lambda) \\
\textrm{Pr}_{NegBin}(y|r,p) &= {k+r-1 \choose r-1} (1-p)^k p^r; \quad r \in \mathbb{N}, \ 0 \! > \! p \! > \! 1.
\end{align}

\begin{figure}[t!]
  \begin{center}
    \includegraphics[width=0.9\textwidth, trim= 0 0.5in 0 1.5in]{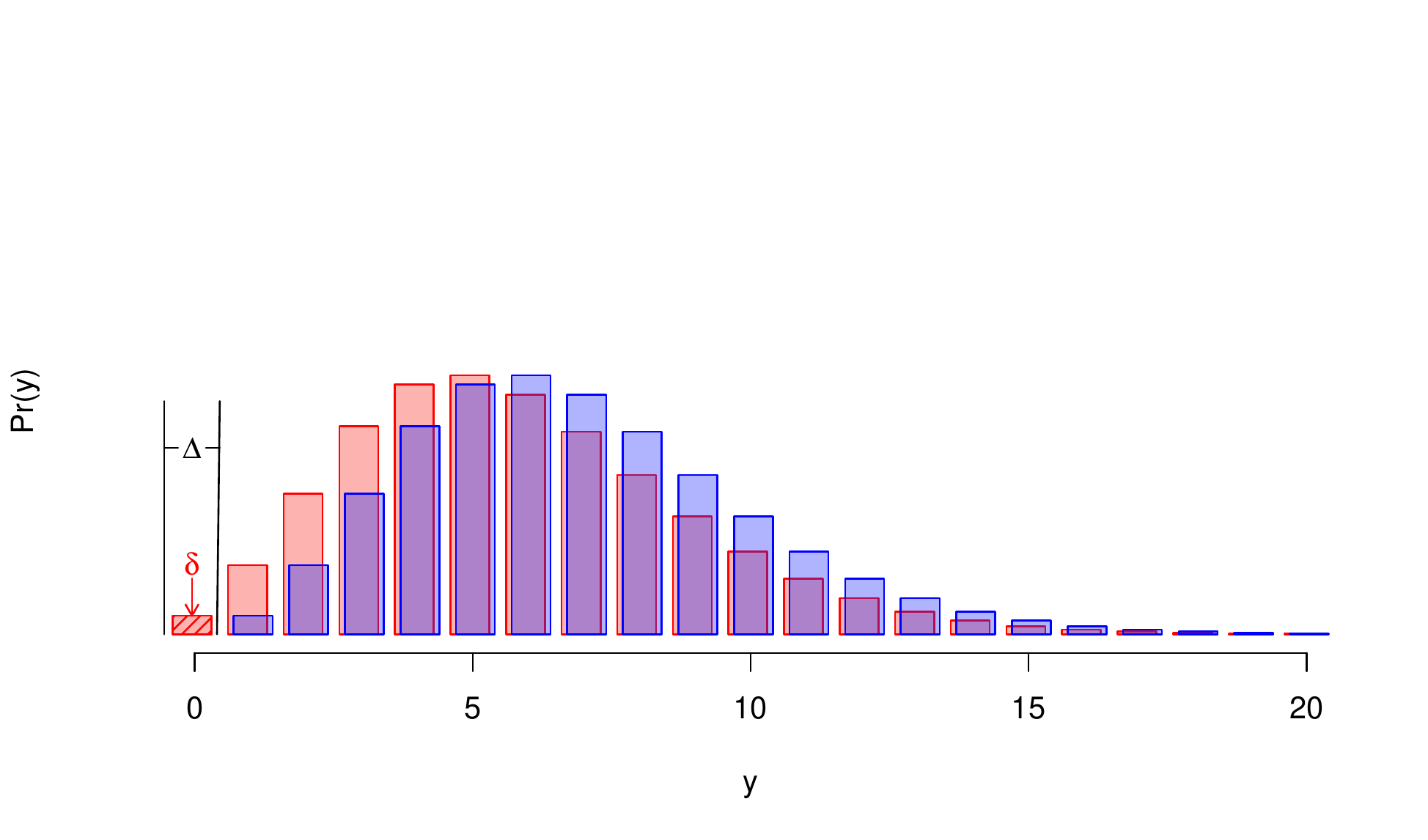}
    \end{center}
   \caption{\em Depicted are two negative binomial distributions with the same parameters, offset by some worst-case sensitivity $\Delta$, where here $\Delta = 1$.  The shaded region (exaggerated for visibility) shows the region with support in only one distribution which has to be covered by $\delta$. }
 \label{f:negbin}
\end{figure}

We first show the ratio of the tails converges:
\begin{align}
\max_{X,X'}\frac{Pr[M(X')=k]}{Pr[M(X)=k]} &= \max_k \frac{ {k+r-1 \choose r-1} (1-p)^k p^r }{{k+r-1+\Delta \choose r-1} (1-p)^{k+\Delta} p^r} 
= \max_k\frac{ (k+r-1)! (k+\Delta)!}{(k+r-1+\Delta)!(k)!} (1-p)^{-\Delta} \nonumber \\
&= (1-p)^{-\Delta} \lim_{k\rightarrow \infty}\frac{\prod_{j=1}^\Delta k+j }{ \prod_{j=1}^\Delta k+r-1+j} 
= (1-p)^{-\Delta}
\end{align}
Which allows us to solve for $\epsilon$ as:
\begin{align}
e^\epsilon \geq (1-p)^{-\Delta} \quad \Rightarrow \quad \epsilon = -\Delta \; \textrm{ln}(1-p)  \quad \Leftrightarrow \quad p = 1-\textrm{exp}(-\frac{\epsilon}{\Delta})   
\end{align}
Note this is roughly $p \approx \epsilon/\Delta, \forall \epsilon/\Delta < 1$. Since $\epsilon$ and $p$ are related, we attempt to fix $\delta$ simply as a function of the remaining free parameter, $r$.  As before:
\begin{align}
\delta \geq \sum_{j=0}^{\Delta-1} Pr_{NegBin}(j|p,r)
\end{align}
For the common case where $\Delta=1$, as for example depicted in Figure \ref{f:negbin}, this means:
\begin{align}
\delta \geq {r-1 \choose r-1} (1-p)^0 p^r = p^r \quad \Rightarrow \quad \delta = p^r \quad \Leftrightarrow \quad r = \Big\lceil \frac{\textrm{ln}(\delta)}{\textrm{ln}(p)} \Big\rceil
\end{align}

\subsection{Example}
Consider again we have $\epsilon=0.5$, and $\delta=10^{-6}$ for a sum or counting query with sensitivity $\Delta=1$.  This gives us $p=0.39$ from which we can compute $r=15$.  The negative binomial has expection $r(1-p)/p$ which calculates to an expected value of 23.2 using the negative binomial mechanism.  This is a close but slightly lower number of padded values than using the truncated Laplace or geometric mechanisms.

%% file: mechanismHeuristics.tex
\subsection{Discretized Uniform and Binomial Distributions}

Two heuristics for padding of records that we have seen are to add records drawn from a uniform distribution, or flip a (typically small) number of coins and add records for every coin that comes up heads.  Since we now have a framework for formalizing the privacy of padding distributions, we briefly point out the associated DP privacy parameters these heuristics imply.  
\begin{alignat}{7}
Pr_{DisUniform}(k|N) &= 1/(N+1); &&\quad k\in \{0, 1, \cdots, N\} &\Rightarrow \epsilon &= 0 & \quad \delta &= \Delta/(N+1)\\
Pr_{Binomial}(k|N) &= {N \choose k} 0.5^N ; &&\quad k\in \{0, 1, \cdots, N\}              &\Rightarrow \epsilon &= \textrm{ln}  {N \choose \Delta}   & \quad \delta &= 0.5^N \sum_{j=0}^{\Delta-1} {N \choose j}  
\end{alignat}
From this perspective we see: (1) the discretized uniform mechanism has a very large $\delta$ term, unless $N$ is of the same order as the size of the dataset, (2) the binomial distribution has a very large $\epsilon$ for any $N$ that gives a conventionally sized $\delta$.

%% file: applicationPSI.tex
\section{Application to Private Set Intersection}
Private Set Intersection considers the problem where two parties $X$ and $Y$ hold private sets $D_X, D_Y \subset D$ and wish to compute some function on the intersection of their two sets; this function could i) reveal the intersection to one or both parties $f(D_X,D_Y) = D_X \cap D_Y$ \cite{meadows1986more,dhoriginal,freedman2004efficient,kissner2005privacy} or ii) compute the cardinality of the intersection or union $f(D_X,D_Y) = |D_X \cap D_Y|$ or $f(D_X,D_Y) = |D_X \cup D_Y|$ \cite{CGT2011psica} or iii) as more recent research on PSI has focused, enable a more general computation on the intersection and associated values to each record and only have parties learn this intended output.  For instance, \cite{ion2019pjc} appends one of the sets $D_X$ with values $(D_X, V_X)$ and computes the sum of values for intersected records $\sum_{x_i = y_j} v_i$ and \cite{movahedi2021privacy} computes a sum also conditioned on a comparison, for $(D_X, T_X, V_X)$ and $(D_Y, T_Y)$ compute $\sum_{x_i = y_j, t_i \geq t_j }v_i$. 

The intended outputs of $f$ can be secured by various DP methods depending on the nature of $f$, e.g. making a sum of associated values differentially private.  However, a common issue that many PSI protocols have is an additional leakage of the intersection size $|D_X\cap D_Y|$ and union size $|D_X \cup D_Y|$ besides the intended outputs. This is the case with the Private-ID, PS3I, and Private Join and Compute protocols \cite{buddhavarapu2020private,ion2019pjc,buddhavarapu2021multikey}.  

We assume a PSI protocol will leak intersection and union sizes, and that any fix should treat the protocol as a black box and be entirely fixed by means of the datasets submitted by the parties. This requires both parties to be semi-honest, which is the assumption already being made in most PSI protocols.  To do this we consider how two parties can each independently draw observations from a shared pool to pad the private data they submit to PSI.  Whenever a collision occurs, that is, both parties draw the same fictitious observation from the pool, then the intersection count is padded by one more unit.  We want to ensure (1) that neither party can work out the other's padded observations (or the collisions) so they can not reverse engineer the noise addition, and (2) the number of collisions---which increase the intersection size leaked by PSI---is guaranteed to form a differentially private distribution and thus sufficiently mask the true intersection size.  To additionally protect the size of the union we will have both parties draw fictitious records from non-intersecting pools and append them to their input sets. 

Our techniques naturally apply to give output privacy to protocols such as \cite{CGT2011psica} that compute the cardinality of the intersection and the union.  Other PSI protocols that our techniques can be applied to include the semi-honest constructions from \cite{PJC2019,buddhavarapu2020private,garimella2021} that compute functions on the intersection while leaking the size of the intersection in the process.  Our techniques are straighforward to apply as long as any associated values to dummy rows can be assigned appropriately so as not the change the intended output.  Since our construction is semi-honest, it is not compatible with the maliciously secure private intersection-sum and cardinality protocols of \cite{miao2020two}.

\subsection{DP intersection size for both parties} \label{sec_int_size}

Let $X$ and $Y$ be two parties who have private finite sets $D_X, D_Y \subset D$ whose intersection is unknown of size $I= | D_X \cap D_Y |$.  Let $A_X$ and $A_Y$ be public finite disjoint sets with no intersection to each other or $D$.  Let $p(\epsilon,\delta)$ be a non-negative differentially private probability distribution on $\mathbb{N}_{0}$, such as from those developed in Section \ref{s:mech}.  We have party $X$ draw an integer $z_x \sim p(\epsilon_x, \delta_x)$ and then sample a random subset $a_X \subset A_X$ of size $z_x$.    They then submit a padded dataset $D_X^+ = \{ D_X \cup a_X \cup A_Y \}$ to some PSI protocol, that is, they combine their private data, with their recent sample and the entirety of the set they did not sample from.  In parallel, $Y$ draws $z_y \sim p(\epsilon_y, \delta_y)$, samples $a_Y \subset A_Y$ of size $z_y$, and submits padded data $D_Y^+ =\{ D_Y \cup A_X \cup a_Y \}$.  What is core to see is that the random subset $X$ generates, $a_X$, will all have collisions in the PSI protocol because $Y$ submits the superset $A_X$, and vice versa.

To ensure that there are $z_x$ or $z_y$ elements available to be sampled, we can use the doubly truncated geometric distribution which has a known maximum which we use to populate $A_X$ and $A_Y$ accordingly (otherwise we need $A_X$, $A_Y$ to be unbounded sets, which even if approximated increases communication).  When we run PSI on $D_X^+$ and $D_Y^+$ we leak the intersection size, which is $I + z_x + z_y$, to both parties.  Party $X$ privately knows $z_x$ so can subtract that and learn $I + z_y$ which is still $(\epsilon_y, \delta_y)$-DP.  Correspondingly, party $Y$ can subtract $z_y$ and only learn $I + z_x$ which is $(\epsilon_x, \delta_x)$-DP.  Thus each party only learns a differentially private answer to the intersection size, whose privacy-loss is controlled by their adversary.

The additional computation and communication is low.  Recall that for the doubly truncated geometric distribution the expectation, $n$, is given in Eq.~\ref{eq:expectN} as a function of $\epsilon$ and $\delta$, and is typically of size 10--100.  For simplicity assume both parties use the same privacy-loss parameters, and thus the same $n$.  Then the sets $A_X$ and $A_Y$ will be of size $2n$, the expected padding to $D_X^+, D_Y^+$ will each be $3n$ and the padding to $\{ D_X^+ \cap D_Y^+ \}$ will have expectation $2n$.

Many PSI protocols that reveal the size of the intersection also reveal the size of the union  $|D_X\cup D_Y|$ as a result of revealing the sizes of the input sets $D_X$ and $D_Y$; this reveals the union size since $|D_X\cup D_Y| = |D_X| +|D_Y| - |D_X\cap D_Y|$.  When we  apply our above technique for DP intersection sizes, it does not give a DP protection to the size of the union, but we can modify it to do so.  This is because the size of the union is revealed when running on our input sets $D_X^+ = \{D_X \cup a_X \cup A_Y\}$ and $D_Y^+ = \{ D_Y \cup A_X \cup a_Y \}$.  The union size is $|D_X^+ \cup D_Y^+| = |D_X \cup D_Y| + 2n$ where $n = |A_X|=|A_Y|$, and since $n$ is public both parties learn the true size of $|D_X \cup D_Y|$.  

The reason this does not also leak the size of the intersection through the relation $|D_Y^+ \cup D_X^+| = |D_X^+| + |D_Y^+| - |D_X^+ \cap D_Y^+|$ is that party $X$ (without loss of generality) in the formula 
\begin{align*}
|D_Y^+ \cup D_X^+| &= |D_X^+| + |D_Y^+| - |D_X^+ \cap D_Y^+|\\
|D_Y \cup D_X| + 2n &= (|D_X| + z_x) + (|D_Y|+z_y)  - (|D_X\cap D_Y| + z_x + z_y)
\end{align*}    
knows the value of the left hand side, and the values of $|D_X|$, $z_X$,  $|D_Y| + z_y$ and $|D_X\cap D_Y|  + z_y$.  So moving all the terms $X$ knows to the left side
\begin{align*}
|D_Y \cup D_X| + 2n - |D_X| &= (|D_Y|+z_y)  - (|D_X\cap D_Y| + z_y)
\end{align*} 
we see $X$ cannot solve for any of the values of $|D_Y|$, $z_y$, or $|D_X \cap D_Y|$ and can only conclude that $|D_Y| + |D_X\cap D_Y| = v$ where $X$ knows the value of $v$.  Said another way $X$ learns only a differentially private size of the party $Y$'s input and the intersection. 
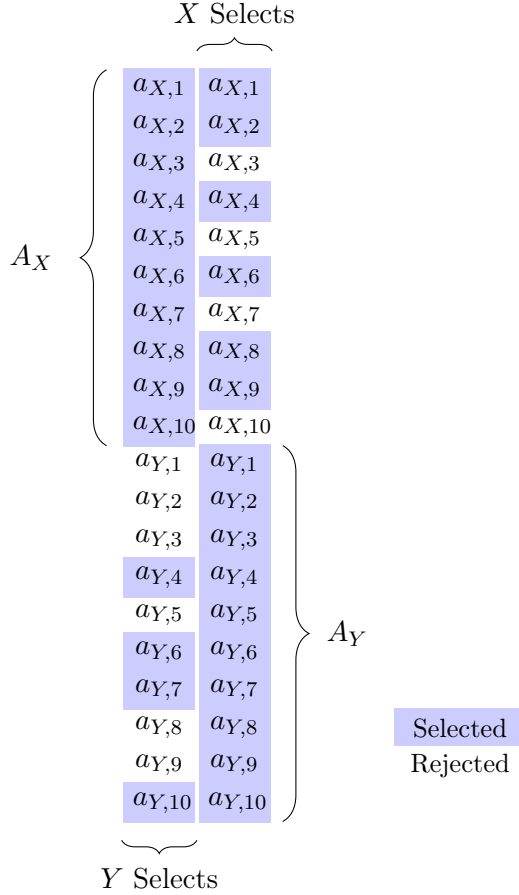
\begin{figure}[h!]
	\centering
	\input{./figs/collision.tikz}
	\caption{\emph{The common pool of observations with example selections from the collision protocol.  Party $X$, whose selections are shown on the right, randomly samples from $A_X$ and takes all of $A_Y$.  Party $Y$, shown on left, randomly samples from $A_Y$ and takes all of $A_X$.  Each party's random samples are guaranteed to collide in the PSI protocol, creating padded users from the desired differentially private distribution.} }
\end{figure}

\subsection{DP union size for both parties} 
We can extend our method to also give differential privacy to union cardinality by considering another two sets $B_X$ and $B_Y$ which are public finite disjoint sets with no intersection to each other or the sets $D$, $A_X$, or $A_Y$.  Similar to before we let $p(\epsilon, \delta)$ be a non-negative differentially private probability distribution on $\mathbb{N}_{0}$ and have party $X$ draw an integer $v_x \sim p(\epsilon_x, \delta_x)$ and then sample a random subset $b_X \subset B_X$ of size $v_x$.  They then submit the padded dataset $D_X^{++} = \{ D_X \cup a_X \cup A_Y \cup b_x \}$.  In parallel $Y$ draws $v_y \sim p(\epsilon_y, \delta_y)$ and samples $b_Y \subset B_Y$ of size $v_x$ and submits the padded data $D_Y^{++} = \{  D_Y \cup A_X \cup a_Y \cup b_Y \}$. Now the additionally padded sets $b_x$ and $b_Y$ will not have any collisions and so will not contribute to the intersection size but they will contribute to the union size so that $|D_X^{++} \cup D_Y^{++}| = |D_X  \cup D_Y| + 2n + v_x + v_y$.  These sets are illustrated in Figure \ref{fig_union} in the Appendix. Thus if a PSI protocol reveals or leaks the size of the union, Party $X$ knows $v_x$ and $n$ and so can subtract to learn $|D_X  \cup D_Y| + v_y$ which is still $(\epsilon_y, \delta_y)$-DP.  Correspondingly, party $Y$ can learn $|D_X  \cup D_Y| + v_y$ which is still $(\epsilon_x, \delta_x)$-DP. Thus each party only learns a differentially private answer to the union size whose privacy-loss is controlled by their adversary.

\subsection{DP intersection size for one party}
In some compute settings, only one party, say $X$, observes the intersection size from PSI.  In this simpler setting we only need to conceal the intersection size from that party.  We can then have $Y$ submit $D_Y^+ = \{D_Y \cup a_Y \}$, where $a_Y$ is generated as before, and $X$ submit $D_X^+ = \{D_X \cup A_Y \}$.  

\subsection{Integration with the Private-ID protocol}
The private Private-ID protocol \cite{buddhavarapu2020private} allows the parties to privately compute a set of pseudorandom universal identifiers (UID) corresponding to the records in the union of their sets, where each party additionally learns which UIDs correspond to which items in its set but not if they belong to the intersection or not. This allows both parties to independently sort their UIDs and the associated records and feed them to any general purpose MPC that ignores the non-matching records and computes on the matching ones. This protocol leaks the sizes of the input sets, union, and intersection to both parties.  Applying our techniques for DP intersection causes the sizes of the input sets and intersection to be differential private.  If we additionally apply our techniques for DP unions, the size of the union becomes differentially private.  In the downstream usage of the UID, both parties will input null associated records for the dummy rows created by the DP noise.  This will allow the DP noise added to secure the leakages of the Private-ID protocol not to effect the actual outcomes computed in any downstream MPC process, which may have its own separate DP mechanisms.
Applying our techniques to the multi-identifier version of the Private-ID protocol \cite{buddhavarapu2021multikey} is a bit more complicated due to the greater leakage in that protocol; we leave this as future work. 
\subsection{Integration with the PS3I protocol}
The PS3I protocol is a decisional Diffie–Hellman (DDH) PSI variant that attaches additive homomorphic encryptions (HE) of associated values to each identifier; the output is secret shares of the associated values for matched records.  Both parties learn the sizes of each others input sets, the intersection size, and thus the union size.  If we apply our technique for DP intersection sizes, we make the sizes of the input sets and the intersection size differentially private, and further if we apply our technique for DP union sizes, we make the union size differentially private.  These additional dummy rows can be assigned a zero associated value or other null value which will be secret shared between the two parties and passed to some other downstream MPC.  

\subsection{Integration with the Private Join and Compute protocol}
The Private Join and Compute protocol \cite{PJC2019} is similar to the PS3I protocol except that attached HE encrypted values are not secret shared but rather joined for users in the intersection.  Downstream operations on these encrypted values are thus limited by the additive HE scheme to being linear operations.  Both parties learn the sizes of each others input sets, the intersection size, and thus the union size.  If we apply our technique for DP intersection sizes, we make the sizes of the input sets and the intersection size differentially private, and further if we apply our technique for DP union sizes, we make the union size differentially private.  The dummy rows can include homomorphic encryption of zero so as not to change the result of the downstream additive HE calculation.  Of course, the additive HE could be replaced with fully homomorphic encryption enabling other additional operations in the downstream, and depending on the nature of the downstream computation other values might be included for the null values. For instance, if the downstream included all multiplications, then encrypting the multiplicative identity, 1, for dummy rows would ensure the downstream result was not effected by the noise.

%% file: figs/collision.tikz.tex
\tikzstyle{select}= [rectangle, fill=blue!20, text centered, text width=7mm,]
\tikzstyle{leave}= [rectangle, text centered, text width=7mm,]
\tikzstyle{legend1}= [rectangle, fill=blue!20, text centered, text width=15mm,]
\tikzstyle{legend2}= [rectangle, text centered, text width=15mm,]

\begin{tikzpicture}

\node[leave] (a0)  at (1,10.5) {};
\node[select] (a1) at (1,10) {$a_{X,1}$} ;
\node[select] (a2) at (1,9.5) {$a_{X,2}$};
\node[select] (a3) at (1,9) {$a_{X,3}$};
\node[select] (a4) at (1,8.5) {$a_{X,4}$};
\node[select] (a5) at (1,8)  {$a_{X,5}$};
\node[select] (a6) at (1,7.5) {$a_{X,6}$};
\node[select] (a7) at (1,7)  {$a_{X,7}$};
\node[select] (a8) at (1,6.5) {$a_{X,8}$};
\node[select] (a9) at (1,6) {$a_{X,9}$};
\node[select] (a10) at (1,5.5) {$a_{X,10}$};
\node[leave] (a11) at (1,5) {$a_{Y,1}$};
\node[leave] (a12) at (1,4.5) {$a_{Y,2}$};
\node[leave] (a13) at (1,4)  {$a_{Y,3}$};
\node[select] (a14) at (1,3.5) {$a_{Y,4}$};
\node[leave] (a15) at (1,3)  {$a_{Y,5}$};
\node[select] (a16) at (1,2.5) {$a_{Y,6}$};
\node[select] (a17) at (1,2) {$a_{Y,7}$};
\node[leave] (a18) at (1,1.5) {$a_{Y,8}$};
\node[leave] (a19) at (1,1) {$a_{Y,9}$};
\node[select] (a20) at (1,0.5) {$a_{Y,10}$};
\node[leave] (a21) at (1,0) {};

\node[leave] (b0)  at (2,10.5) {};
\node[select] (b1) at (2,10) {$a_{X,1}$} ;
\node[select] (b2) at (2,9.5) {$a_{X,2}$};
\node[leave] (b3) at (2,9) {$a_{X,3}$};
\node[select] (b4) at (2,8.5) {$a_{X,4}$};
\node[leave] (b5) at (2,8)  {$a_{X,5}$};
\node[select] (b6) at (2,7.5) {$a_{X,6}$};
\node[leave] (b7) at (2,7)  {$a_{X,7}$};
\node[select] (b8) at (2,6.5) {$a_{X,8}$};
\node[select] (b9) at (2,6) {$a_{X,9}$};
\node[leave] (b10) at (2,5.5) {$a_{X,10}$};
\node[select] (b11) at (2,5) {$a_{Y,1}$};
\node[select] (b12) at (2,4.5) {$a_{Y,2}$};
\node[select] (b13) at (2,4)  {$a_{Y,3}$};
\node[select] (b14) at (2,3.5) {$a_{Y,4}$};
\node[select] (b15) at (2,3)  {$a_{Y,5}$};
\node[select] (b16) at (2,2.5) {$a_{Y,6}$};
\node[select] (b17) at (2,2) {$a_{Y,7}$};
\node[select] (b18) at (2,1.5) {$a_{Y,8}$};
\node[select] (b19) at (2,1) {$a_{Y,9}$};
\node[select] (b20) at (2,0.5) {$a_{Y,10}$};
\node[leave] (b21) at (2,0) {};

\node[legend1] (l1) at (5,1.5) {\small{Selected}};
\node[legend2] (l2) at (5,1) {\small{Rejected}};


\draw [decorate,decoration={brace,amplitude=10pt,mirror}]
($(a0)!0.5!(a1) - (0.7,0)$) -- ($(a10)!0.5!(a11) - (0.7,0)$) node [black,midway,xshift=-1cm] 
{$A_X$};
\draw [decorate,decoration={brace,amplitude=10pt,raise=4pt}]
($(b10)!0.5!(b11) + (0.5,0)$) -- ($(b20)!0.5!(b21) + (0.5,0)$) node [black,midway,xshift=1cm] {$A_Y$};

\draw [decorate,decoration={brace,amplitude=5pt}]
($(b0) -(0.5,0)$) -- ($(b0) + (0.5,0)$) node [black,midway,yshift=0.5cm] 
{$X$ Selects};

\draw [decorate,decoration={brace,amplitude=5pt}]
($(a21) + (0.5,0)$) -- ($(a21) - (0.5,0)$) node [black,midway,yshift=-0.5cm] 
{$Y$ Selects};

\end{tikzpicture}

%% file: applicationDPHist.tex
\section{Application to MPC side-channel attacks and DP Histograms}

The blending of secure multiparty computation and differential privacy (MPDPC) as interwoven privacy enhancing technologies has the promise to offer privacy across stores, across computation, and across releases, that is, enabling a computation to occur across the data of multiple parties while encrypting the data during computation and offering differentially private answers at the conclusion (sometimes described as input and output privacy respectively).  In simple settings this blend means the algorithm that is encoded in the MPC game needs itself to have differentially private outputs, such as noise mechanisms for the released values.  However, in the typical DP threat models (particularly in the centralized curator model) the ``output'' is only the final answers released to the world at the end of the computation, whereas in the MPC threat model, the act of computation is itself a continuous output under observation by the adversarial parties.  While MPC shields all the data values and intermediate calculations, it is often susceptible to inference on the data by sidechannels, a risk receiving increasing attention.

Consider an MPC implementation that counts a grand sum of a predicate over user records.  This is a common generic task, as for example in the style of \cite{movahedi2021privacy}, where a dataset consists of the events associated with each user, and we are counting the sum of all user events that meet a filter.  The grand sum can be made differentially private by noise additon, however, oftentimes one loops first over users, and then over those user's events.  Within the MPC calculation either (1) the storage access pattern, or (2) the total time to compute the predicate over that user's records, can directly leak the number of events for each user.

Thus while MPC encrypts the data, storage access and timing present common side channels, which are themselves an output (as for example explicitly considered by \cite{haeberlen2011differential}).  If the goal of MPDPC is to privatize all outputs, then we need DP promises on all such channels.  The main solution to timing attacks and storage in MPC has been to enforce constant-time/constant-storage computations, which typically entail lengthening compute time to the worst case, often at severe efficiency loss.  

\subsection{Non-negative padding for side-channel solution}
A partial solution to this is to shuffle the order in which users are evaluated.  This means if we leak compute time or record recall for the first user in the dataset and determine they had five events, we don't know which user had those five events.  However, at the end of the grand sum, we will have witnessed the number of events for each user in the dataset, leaking the histogram of the number of events.  If an adversary can rerun the computation with a differencing attack, say by removing Alice's data, then the category in the histogram that changed will reveal Alice's data, deterministically.

The constant-time/constant-storage solution to this is to pad all users with the maximum number of events (but make sure that the fake events fail the predicate being counted).  From the differencing attack perspective, now the histogram has only one category and no information is revealed.  However, the computational cost of this can be enormous, particularly if a rare number of important users have very many events.

Instead of padding each user's events, non-negative DP noise can be used to generate new users whose data makes the leaked histogram differentially private.   Consider a dataset, $D$, of $N$ individuals who can have up to $\#d_i \leq K$ events, and assume the histogram of $\#d$ is leaked by timing.  For each $i=\{0,\cdots,K\}$ we draw $j_i \sim p(\epsilon,\delta)$ from a non-negative DP noise distribution on $\mathbb{N}_0$, and add $j_i$ users with $i$ (predicate failing) events to dataset $D$.  The resulting histogram is $(\epsilon, \delta)$-DP and masks the histogram of user events.

\subsection{Example}
Assume users are roughly uniformly distributed in number of events across $\{0, \cdots, K\}$.  This is conservative as oftentimes such distributions have modes at zero and heavy tails.  A constant-time solution requires each user now to have exactly $K$ events for MPC to cycle over, meaning we have padded the data with $N \!  K/2$ events in expectation.  If instead we add DP padded users to each event number, we add $\sum_{i=0}^K n K = nK(K+1)/2$ padded events, where $n$ is again the expectation of the DP distribution $p$ as in Eq.~\ref{eq:expectN}.  For datasets with many more users than possible events, that is $N \! \gg \! K$, then $n(K \! + \! 1) \! \ll \! N$ and the number of padded events---and hence added MPC compute time---is much lower from using DP non-negative padded users, than from constant-time computation.\footnote{We note that this is not a complete analysis of the timing difference.  There might be costs in shuffling the data, which are not required in the constant-time solution, as well as fixed timing costs in overhead per user (unrelated to number of events).  These can be added for an exact comparison if the timing difference is close, however, the key logic remains.}

\subsection{Shuffle of Private-ID Universal Identifiers}
The Private-ID protocol \cite{buddhavarapu2020private} allows the parties to privately compute a set of pseudorandom universal identifiers (UID) corresponding to the records in the union of their sets, where each party additionally learns which UIDs correspond to which items in its set but not if they belong to the intersection or not. This allows both parties to independently sort their UIDs and the associated records and feed them to any general purpose MPC that ignores the non-matching records and computes on the matching ones.  The number of associated records to be passed into the downstream MPC computation varies. Non-matched records have zero associated records while matched records may have many records depending on the application.  In order to not reveal the number of records per row, we can apply our padding for DP Histograms as long as we can perform a shuffle on the Private-ID UIDs.  The cost of an oblivious shuffle of $n$ elements in MPC is often estimated as requiring $c n log(n)^2$ where $c$ is a constant in the range $[4,7]$.  

\begin{figure}[!ht]
\centering
    \subfloat[Constant time versus DP padded datasets\label{subfig-1:dummy}]{%
      \includegraphics[width=0.45\textwidth]{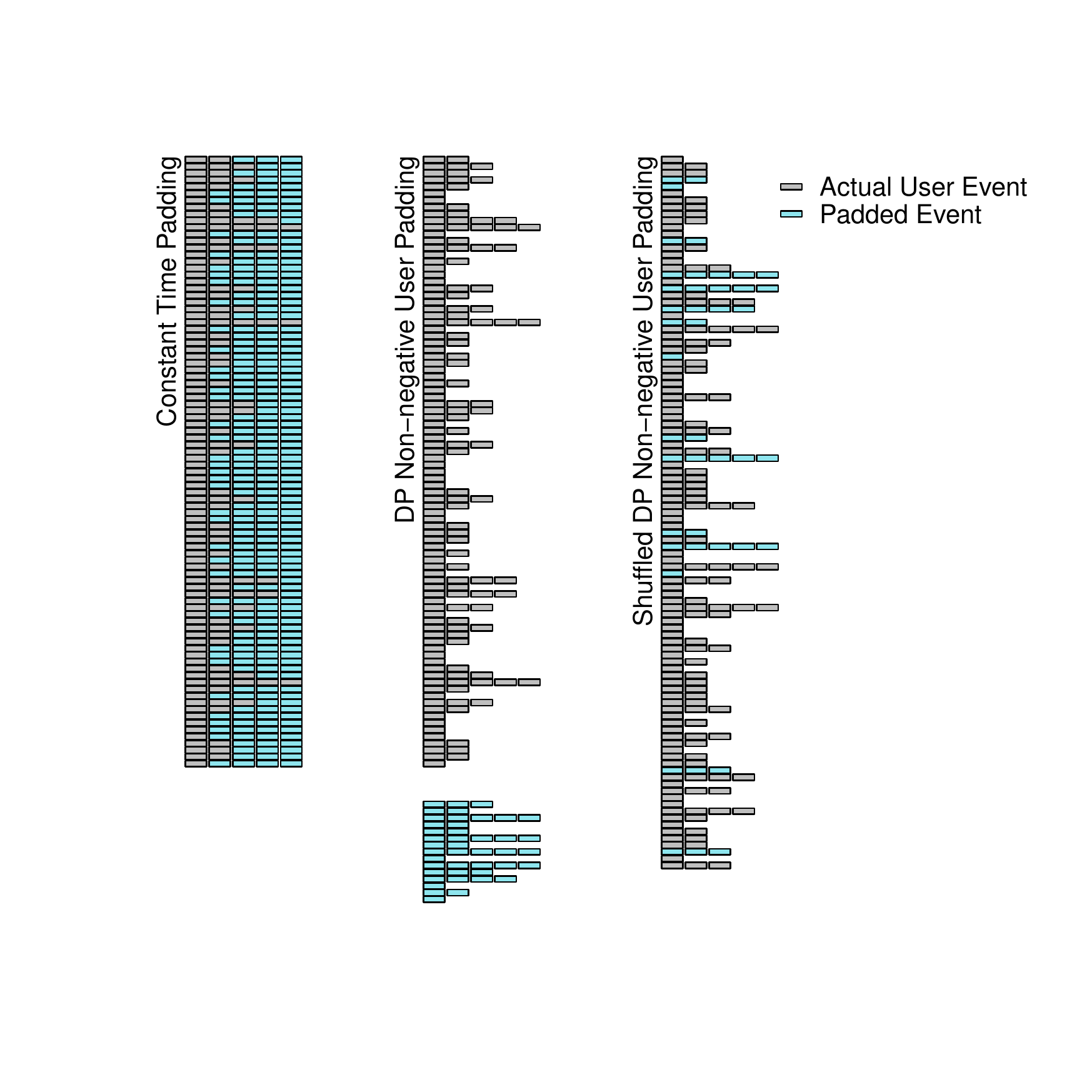}
    }
    \subfloat[Privacy-preserving histogram from DP padding\label{subfig-2:dummy}]{%
      \includegraphics[width=0.45\textwidth]{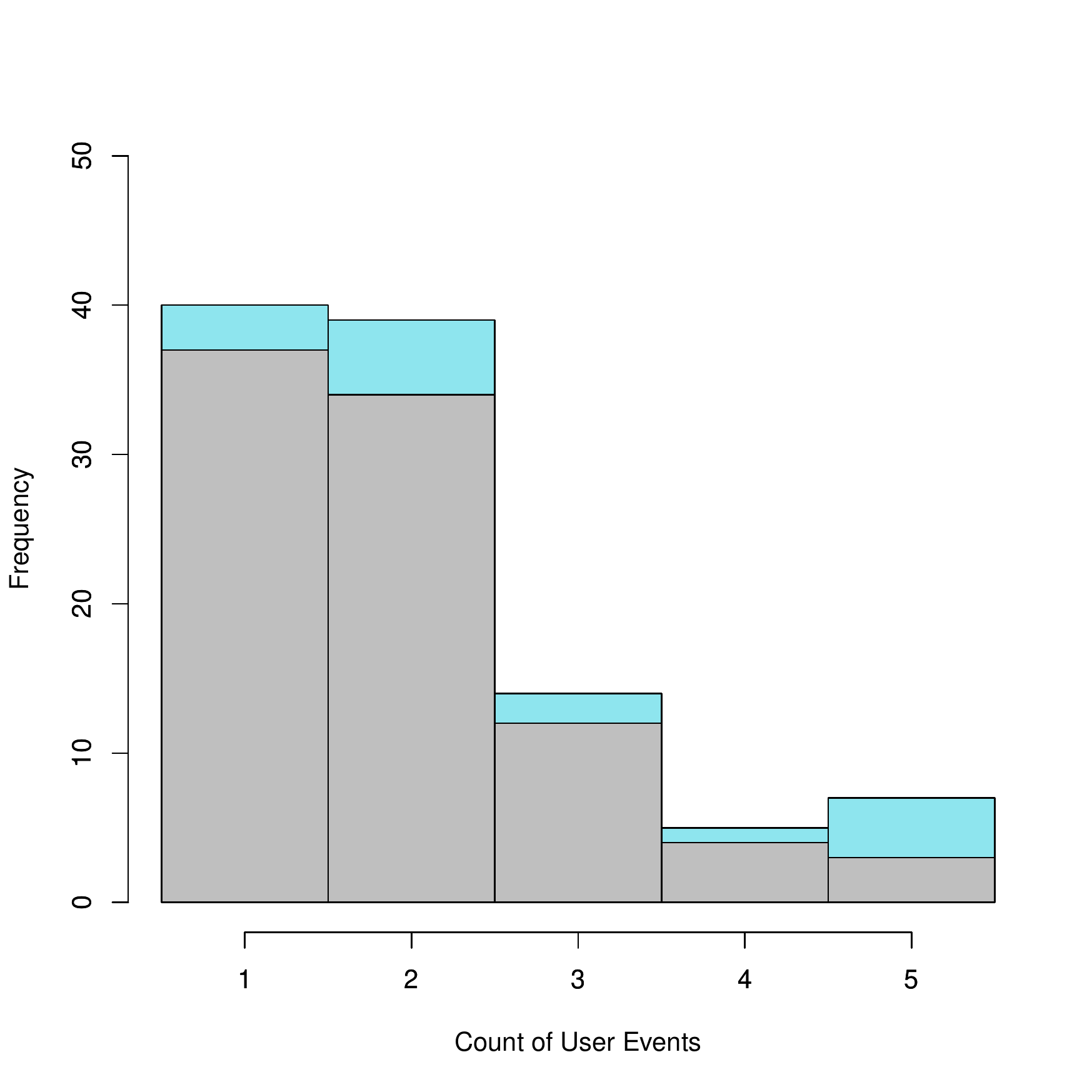}
    }
    \caption{ \emph{On the left figure we see a dataset padding every user to the worst case.  In the center is a dataset plus the padded observations created by adding a DP non-negative number of users for each event count.  These are separated in the center for visual clarity, but are shuffled along with the original data on the right.  Far fewer padded events (blue items) are required with DP padding.\\
    On the right figure we see the privacy preserving histogram that would result from this solution, broken into the contributions by the original data (gray) and padded data (blue).  The resulting blue noise in this DP histogram defeats any differencing attack.}}
    \label{fig:dummy}
  \end{figure}

%% file: priorwork.tex
\section{Prior Work}

The combination of padded records and differential privacy in the context of storage and joins has been recognized as a solution for side-channel attacks.  While we believe our mechanisms are lightweight and straightforward in construction, and thus well tailored to  individual queries, previous papers have achieved positive padding for database stores in large scale ways that permit repeated queries.  Encryted storage techniques, such as Oblivious RAM, have been shown to be susceptible to side-channel attacks on data access patterns and communication volume \cite{kellaris2016generic, grubbs2018pump, gui2019encrypted}.  Bogatov et al.~\cite{bogatov2021mathcaletextpsolute} considers a DP sanitization solution that uses a DP tree hierarchy.  This effectively results in buckets of storage each of which have some padded users and combine to offer a DP guarantee to queries on the store.  Relatedly, Xu et al.~\cite{xu2019hermetic} implement a DP padding solution for the same problem that uses a shifted and tail-censored Laplace.  This is the closest to anything we present, however, their use of a $\delta$ parameter is censoring of the entire left tail of the Laplace, whereas we use $\delta$ for a narrower purpose.  Groce et al.~\cite{groce2019cheaper} also improve this Laplace approach within a Bayesian context.  Allen et al.~\cite{allen2019algorithmic} use a padding procedure on Oblivious RAM that adds sufficiently large numbers of fake records that the standard (zero-centered) ($\epsilon$,0)-DP Laplace mechanism computationally is ensured to result in answers that can be filled with all the true records plus some number of the fake records which act as padding, functionally quite similar to this shifted and tail-censored Laplace mechanism reoccuring in the literature.

Differential privacy has been considered in the context of solving side-channel leakage in private record linkage by \cite{inan2010private,kuzu2013efficient,cao2015hybrid}.  Within private record linkage, He et al.~\cite{he2017composing} explicitly connect this to private set intersection, using the shifted tail-censored Laplace, and \cite{groce2019cheaper} extends this work.  The side-channel they both consider is load estimates in blocks from hashing.

Any DP noise distribution on $\mathbb{R}$ (or $\mathbb{Z}$) can be converted to a non-negative distribution on $\mathbb{R}_{\geq 0}$ (or $\mathbb{N}_0$) by proceedures we have used for truncating and shifting the Laplace and geometric mechanisms.  However, some mechanisms from prior work naturally generate releases on a bounded interval, and like our use of the negative binomial, these could be more readily converted.  Quick \cite{quick2021generating, quick2021comparison, quick2021improving} has a differentially private formulation of the Poisson-gamma distribution that is used to make all DP released counts strictly non-negative, and it would be straightforward to convert this mechanism from making the release non-negative to the noise to be non-negative.  Similarly, implementations of the Dirichlet distribution going back to Machanavajjhala et al.~\cite{machanavajjhala2008privacy} could be so converted.

%% file: conclusion.tex
\section{Conclusion}

Differentially private noise mechanisms typically have symmetric noise about the true sample value.  However, there are useful applications where we require an error with a known sign.  Truncations of the Laplace and geometric distributions, as well as count distributions with subexponential tails, such as the negative binomial, can be converted into non-negative (or non-positive) noise mechanisms under approximate differential privacy.  Such mechanisms are particularly appealing for operations where we need to pad the underlying data with extra observations so as to make any system leakage differentially private, as in oblivious storage, private set intersection, and timing and storage side-channels in MPC.

%% file: appendix.tex
\section{Appendix}\label{a:work}
The derivation of the $\delta$ constraint for the geometric mechanism can be shown as:
\begin{align}
Ae^{-n \epsilon} = Ar^n &= \frac{r^n(1-r)}{1 + r -2r^{n+1}} < \delta\\
 \Rightarrow \quad r^n &< \frac{\delta ( 1 + r -2r^{n+1})}{1-r}\\
\Rightarrow \quad r^n \Big( 1 + \frac{2r\delta}{1-r}\Big) &< \frac{ \delta(1+r)}{1-r}\\
\Rightarrow \quad r^n \Big(\frac{1-r + 2r\delta}{1-r}\Big) &< \frac{ \delta(1+r)}{1-r}\\
\Rightarrow \quad r^n (1-r + 2r\delta) &> \delta (1 + r) \\
\Rightarrow \quad r^n &> \frac{\delta (1 + r)}{1-r + 2r\delta} \\
\Rightarrow \quad e^{-\epsilon n} &> \frac{\delta (1 + r)}{1-r + 2r\delta} \\
\Rightarrow \quad n &= \Big\lceil -\epsilon^{-1} ln \Big(\frac{\delta (1 + r)}{1-r + 2r\delta}\Big) \Big\rceil
\end{align}

\begin{figure}[h!] 	
	\centering
	\input{./figs/union.collision.tikz}
	\caption{\emph{Union and intersection DP padded inputs} }  \label{fig_union}
\end{figure}
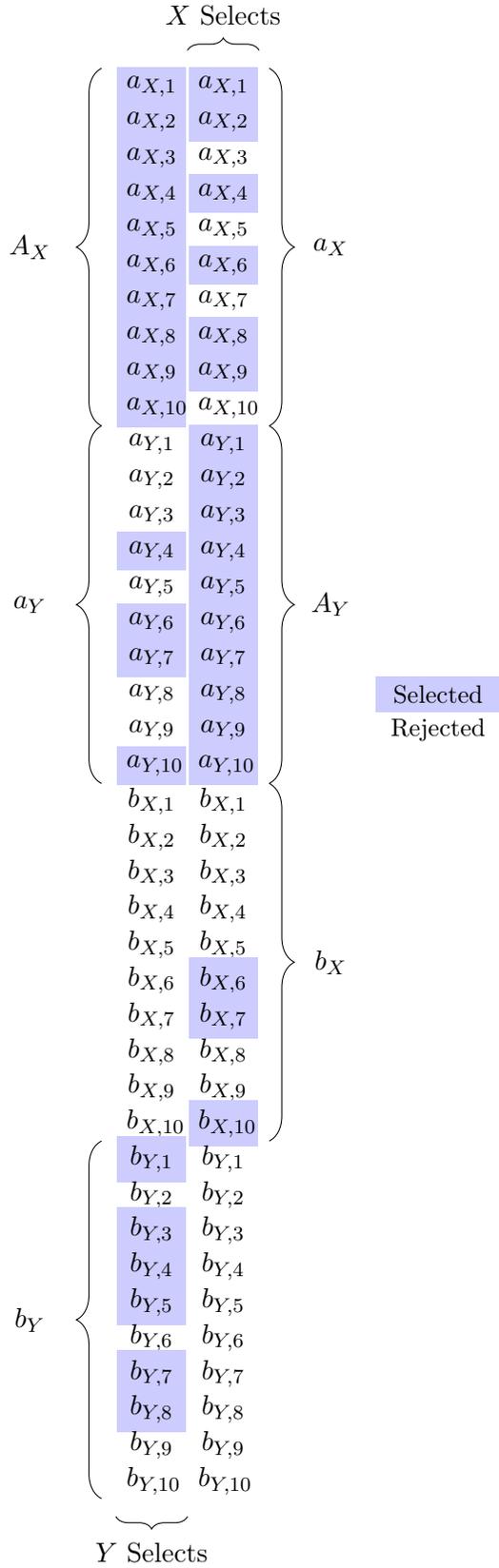

%% file: figs/union.collision.tikz.tex
\tikzstyle{select}= [rectangle, fill=blue!20, text centered, text width=7mm,]
\tikzstyle{leave}= [rectangle, text centered, text width=7mm,]
\tikzstyle{legend1}= [rectangle, fill=blue!20, text centered, text width=15mm,]
\tikzstyle{legend2}= [rectangle, text centered, text width=15mm,]

\begin{tikzpicture}

\node[leave] (a0)  at (1,10.5) {};
\node[select] (a1) at (1,10) {$a_{X,1}$} ;
\node[select] (a2) at (1,9.5) {$a_{X,2}$};
\node[select] (a3) at (1,9) {$a_{X,3}$};
\node[select] (a4) at (1,8.5) {$a_{X,4}$};
\node[select] (a5) at (1,8)  {$a_{X,5}$};
\node[select] (a6) at (1,7.5) {$a_{X,6}$};
\node[select] (a7) at (1,7)  {$a_{X,7}$};
\node[select] (a8) at (1,6.5) {$a_{X,8}$};
\node[select] (a9) at (1,6) {$a_{X,9}$};
\node[select] (a10) at (1,5.5) {$a_{X,10}$};
\node[leave] (a11) at (1,5) {$a_{Y,1}$};
\node[leave] (a12) at (1,4.5) {$a_{Y,2}$};
\node[leave] (a13) at (1,4)  {$a_{Y,3}$};
\node[select] (a14) at (1,3.5) {$a_{Y,4}$};
\node[leave] (a15) at (1,3)  {$a_{Y,5}$};
\node[select] (a16) at (1,2.5) {$a_{Y,6}$};
\node[select] (a17) at (1,2) {$a_{Y,7}$};
\node[leave] (a18) at (1,1.5) {$a_{Y,8}$};
\node[leave] (a19) at (1,1) {$a_{Y,9}$};
\node[select] (a20) at (1,0.5) {$a_{Y,10}$};
\node[leave] (a21) at (1,0) {$b_{X,1}$};
\node[leave] (a22) at (1,-.5) {$b_{X,2}$};
\node[leave] (a23) at (1,-1)  {$b_{X,3}$};
\node[leave] (a24) at (1,-1.5) {$b_{X,4}$};
\node[leave] (a25) at (1,-2)  {$b_{X,5}$};
\node[leave] (a26) at (1,-2.5) {$b_{X,6}$};
\node[leave] (a27) at (1,-3) {$b_{X,7}$};
\node[leave] (a28) at (1,-3.5) {$b_{X,8}$};
\node[leave] (a29) at (1,-4) {$b_{X,9}$};
\node[leave] (a30) at (1,-4.5) {$b_{X,10}$};
\node[select] (a31) at (1,-5) {$b_{Y,1}$};
\node[leave] (a32) at (1,-5.5) {$b_{Y,2}$};
\node[select] (a33) at (1,-6)  {$b_{Y,3}$};
\node[select] (a34) at (1,-6.5) {$b_{Y,4}$};
\node[select] (a35) at (1,-7)  {$b_{Y,5}$};
\node[leave] (a36) at (1,-7.5) {$b_{Y,6}$};
\node[select] (a37) at (1,-8) {$b_{Y,7}$};
\node[select] (a38) at (1,-8.5) {$b_{Y,8}$};
\node[leave] (a39) at (1,-9) {$b_{Y,9}$};
\node[leave] (a40) at (1,-9.5) {$b_{Y,10}$};
\node[leave] (a41) at (1,-10) {};

\node[leave] (b0)  at (2,10.5) {};
\node[select] (b1) at (2,10) {$a_{X,1}$} ;
\node[select] (b2) at (2,9.5) {$a_{X,2}$};
\node[leave] (b3) at (2,9) {$a_{X,3}$};
\node[select] (b4) at (2,8.5) {$a_{X,4}$};
\node[leave] (b5) at (2,8)  {$a_{X,5}$};
\node[select] (b6) at (2,7.5) {$a_{X,6}$};
\node[leave] (b7) at (2,7)  {$a_{X,7}$};
\node[select] (b8) at (2,6.5) {$a_{X,8}$};
\node[select] (b9) at (2,6) {$a_{X,9}$};
\node[leave] (b10) at (2,5.5) {$a_{X,10}$};
\node[select] (b11) at (2,5) {$a_{Y,1}$};
\node[select] (b12) at (2,4.5) {$a_{Y,2}$};
\node[select] (b13) at (2,4)  {$a_{Y,3}$};
\node[select] (b14) at (2,3.5) {$a_{Y,4}$};
\node[select] (b15) at (2,3)  {$a_{Y,5}$};
\node[select] (b16) at (2,2.5) {$a_{Y,6}$};
\node[select] (b17) at (2,2) {$a_{Y,7}$};
\node[select] (b18) at (2,1.5) {$a_{Y,8}$};
\node[select] (b19) at (2,1) {$a_{Y,9}$};
\node[select] (b20) at (2,0.5) {$a_{Y,10}$};
\node[leave] (b21) at (2,0) {$b_{X,1}$};
\node[leave] (b22) at (2,-.5) {$b_{X,2}$};
\node[leave] (b23) at (2,-1)  {$b_{X,3}$};
\node[leave] (b24) at (2,-1.5) {$b_{X,4}$};
\node[leave] (b25) at (2,-2)  {$b_{X,5}$};
\node[select] (b26) at (2,-2.5) {$b_{X,6}$};
\node[select] (b27) at (2,-3) {$b_{X,7}$};
\node[leave] (b28) at (2,-3.5) {$b_{X,8}$};
\node[leave] (b29) at (2,-4) {$b_{X,9}$};
\node[select] (b30) at (2,-4.5) {$b_{X,10}$};
\node[leave] (b31) at (2,-5) {$b_{Y,1}$};
\node[leave] (b32) at (2,-5.5) {$b_{Y,2}$};
\node[leave] (b33) at (2,-6)  {$b_{Y,3}$};
\node[leave] (b34) at (2,-6.5) {$b_{Y,4}$};
\node[leave] (b35) at (2,-7)  {$b_{Y,5}$};
\node[leave] (b36) at (2,-7.5) {$b_{Y,6}$};
\node[leave] (b37) at (2,-8) {$b_{Y,7}$};
\node[leave] (b38) at (2,-8.5) {$b_{Y,8}$};
\node[leave] (b39) at (2,-9) {$b_{Y,9}$};
\node[leave] (b40) at (2,-9.5) {$b_{Y,10}$};
\node[leave] (b41) at (2,-10) {};

\node[legend1] (l1) at (5,1.5) {\small{Selected}};
\node[legend2] (l2) at (5,1) {\small{Rejected}};


\draw [decorate,decoration={brace,amplitude=10pt,mirror}]
($(a0)!0.5!(a1) - (0.7,0)$) -- ($(a10)!0.5!(a11) - (0.7,0)$) node [black,midway,xshift=-1cm] 
{$A_X$};

\draw [decorate,decoration={brace,amplitude=10pt,raise=4pt}]
($(b10)!0.5!(b11) + (0.5,0)$) -- ($(b20)!0.5!(b21) + (0.5,0)$) node [black,midway,xshift=1cm] {$A_Y$};

\draw [decorate,decoration={brace,amplitude=5pt}]
($(b0) -(0.5,0)$) -- ($(b0) + (0.5,0)$) node [black,midway,yshift=0.5cm] 
{$X$ Selects};

\draw [decorate,decoration={brace,amplitude=5pt}]
($(a41) + (0.5,0)$) -- ($(a41) - (0.5,0)$) node [black,midway,yshift=-0.5cm] 
{$Y$ Selects};

\draw [decorate,decoration={brace,amplitude=10pt,mirror}]
($(a10)!0.5!(a11) - (0.7,0)$) -- ($(a20)!0.5!(a21) - (0.7,0)$) node [black,midway,xshift=-1cm] 
{$a_Y$};

\draw [decorate,decoration={brace,amplitude=10pt,raise=4pt}]
($(b0)!0.5!(b1) + (0.5,0)$) -- ($(b10)!0.5!(b11) + (0.5,0)$) node [black,midway,xshift=1cm] {$a_X$};

\draw [decorate,decoration={brace,amplitude=10pt,mirror}]
($(a30)!0.5!(a31) - (0.7,0)$) -- ($(a40)!0.5!(a41) - (0.7,0)$) node [black,midway,xshift=-1cm] 
{$b_Y$};

\draw [decorate,decoration={brace,amplitude=10pt,raise=4pt}]
($(b20)!0.5!(b21) + (0.5,0)$) -- ($(b30)!0.5!(b31) + (0.5,0)$) node [black,midway,xshift=1cm] {$b_X$};

\end{tikzpicture}

%% file: main.bbl
\begin{thebibliography}{10}
\providecommand{\url}[1]{#1}
\csname url@samestyle\endcsname
\providecommand{\newblock}{\relax}
\providecommand{\bibinfo}[2]{#2}
\providecommand{\BIBentrySTDinterwordspacing}{\spaceskip=0pt\relax}
\providecommand{\BIBentryALTinterwordstretchfactor}{4}
\providecommand{\BIBentryALTinterwordspacing}{\spaceskip=\fontdimen2\font plus
\BIBentryALTinterwordstretchfactor\fontdimen3\font minus
  \fontdimen4\font\relax}
\providecommand{\BIBforeignlanguage}[2]{{%
\expandafter\ifx\csname l@#1\endcsname\relax
\typeout{** WARNING: IEEEtran.bst: No hyphenation pattern has been}%
\typeout{** loaded for the language `#1'. Using the pattern for}%
\typeout{** the default language instead.}%
\else
\language=\csname l@#1\endcsname
\fi
#2}}
\providecommand{\BIBdecl}{\relax}
\BIBdecl

\bibitem{kellaris2017accessing}
G.~Kellaris, G.~Kollios, K.~Nissim, and A.~O'Neill, ``Accessing data while
  preserving privacy,'' \emph{arXiv preprint arXiv:1706.01552}, 2017.

\bibitem{PJC2019}
M.~Ion, B.~Kreuter, A.~E. Nergiz, S.~Patel, M.~Raykova, S.~Saxena, K.~Seth,
  D.~Shanahan, and M.~Yung, ``On deploying secure computing: Private
  intersection-sum-with-cardinality,'' Cryptology ePrint Archive, Report
  2019/723, 2019, \url{https://ia.cr/2019/723}.

\bibitem{buddhavarapu2020private}
P.~Buddhavarapu, A.~Knox, P.~Mohassel, S.~Sengupta, E.~Taubeneck, and
  V.~Vlaskin, ``Private matching for compute.'' \emph{IACR Cryptol. ePrint
  Arch.}, vol. 2020, p. 599, 2020.

\bibitem{movahedi2021privacy}
M.~Movahedi, B.~M. Case, J.~Honaker, A.~Knox, L.~Li, Y.~P. Li, S.~Saravanan,
  S.~Sengupta, and E.~Taubeneck, ``Privacy-preserving randomized controlled
  trials: A protocol for industry scale deployment,'' \emph{arXiv preprint},
  2021.

\bibitem{haeberlen2011differential}
A.~Haeberlen, B.~C. Pierce, and A.~Narayan, ``Differential privacy under
  fire.'' in \emph{USENIX Security Symposium}, vol.~33, 2011.

\bibitem{covington2020unknown}
C.~Covington, J.~Honaker, and M.~Shoemate, ``Smartnoise: Working with unknown
  dataset sizes,'' Notebook, 2020,
  \url{https://github.com/opendp/smartnoise-samples/blob/master/analysis/unknown_dataset_size.ipynb}.

\bibitem{Kotz2001}
S.~Kotz, T.~Kozubowski, and K.~Podgorski, \emph{The Laplace Distribution and
  Generalizations: A Revisit with Applications to Communications, Economics,
  Engineering, and Finance}.\hskip 1em plus 0.5em minus 0.4em\relax Boston:
  Birkh{\"a}user Basel, 2001.

\bibitem{CGT2011psica}
E.~De~Cristofaro, P.~Gasti, and G.~Tsudik, ``Fast and private computation of
  cardinality of set intersection and union,'' in \emph{International
  Conference on Cryptology and Network Security}.\hskip 1em plus 0.5em minus
  0.4em\relax Springer, 2012, pp. 218--231.

\bibitem{ion2019pjc}
M.~Ion, B.~Kreuter, A.~E. Nergiz, S.~Patel, M.~Raykova, S.~Saxena, K.~Seth,
  D.~Shanahan, and M.~Yung, ``On deploying secure computing: Private
  intersection-sum-with-cardinality,'' Cryptology ePrint Archive, Report
  2019/723, 2019, \url{https://eprint.iacr.org/2019/723}.

\bibitem{buddhavarapu2021multikey}
P.~Buddhavarapu, B.~M. Case, L.~Gore, A.~Knox, P.~Mohassel, S.~Sengupta,
  E.~Taubeneck, and M.~Xue, ``Multi-key private matching for compute,''
  Cryptology ePrint Archive, Report 2021/770, 2021,
  \url{https://ia.cr/2021/770}.

\bibitem{garimella2021}
G.~Garimella, P.~Mohassel, M.~Rosulek, S.~Sadeghian, and J.~Singh, ``Private
  set operations from oblivious switching,'' Cryptology ePrint Archive, Report
  2021/243, 2021, \url{https://ia.cr/2021/243}.

\bibitem{miao2020two}
P.~Miao, S.~Patel, M.~Raykova, K.~Seth, and M.~Yung, ``Two-sided malicious
  security for private intersection-sum with cardinality,'' in \emph{Annual
  International Cryptology Conference}.\hskip 1em plus 0.5em minus 0.4em\relax
  Springer, 2020, pp. 3--33.

\bibitem{kellaris2016generic}
G.~Kellaris, G.~Kollios, K.~Nissim, and A.~O'Neill, ``Generic attacks on secure
  outsourced databases,'' in \emph{Proceedings of the 2016 ACM SIGSAC
  Conference on Computer and Communications Security}, 2016, pp. 1329--1340.

\bibitem{grubbs2018pump}
P.~Grubbs, M.-S. Lacharit{\'e}, B.~Minaud, and K.~G. Paterson, ``Pump up the
  volume: Practical database reconstruction from volume leakage on range
  queries,'' in \emph{Proceedings of the 2018 ACM SIGSAC Conference on Computer
  and Communications Security}, 2018, pp. 315--331.

\bibitem{gui2019encrypted}
Z.~Gui, O.~Johnson, and B.~Warinschi, ``Encrypted databases: New volume attacks
  against range queries,'' in \emph{Proceedings of the 2019 ACM SIGSAC
  Conference on Computer and Communications Security}, 2019, pp. 361--378.

\bibitem{bogatov2021mathcaletextpsolute}
D.~Bogatov, G.~Kellaris, G.~Kollios, K.~Nissim, and A.~O'Neill,
  ``$\mathcal{E}\text{psolute}$: Efficiently querying databases while providing
  differential privacy,'' 2021.

\bibitem{xu2019hermetic}
M.~Xu, A.~Papadimitriou, A.~Haeberlen, and A.~Feldman, ``Hermetic:
  Privacy-preserving distributed analytics without (most) side channels,''
  \emph{External Links: Link Cited by}, 2019.

\bibitem{groce2019cheaper}
A.~Groce, P.~Rindal, and M.~Rosulek, ``Cheaper private set intersection via
  differentially private leakage,'' \emph{Proceedings on Privacy Enhancing
  Technologies}, vol. 2019, no.~3, 2019.

\bibitem{allen2019algorithmic}
J.~Allen, B.~Ding, J.~Kulkarni, H.~Nori, O.~Ohrimenko, and S.~Yekhanin, ``An
  algorithmic framework for differentially private data analysis on trusted
  processors,'' \emph{Advances in Neural Information Processing Systems},
  vol.~32, pp. 13\,657--13\,668, 2019.

\bibitem{inan2010private}
A.~Inan, M.~Kantarcioglu, G.~Ghinita, and E.~Bertino, ``Private record matching
  using differential privacy,'' in \emph{Proceedings of the 13th International
  Conference on Extending Database Technology}, 2010, pp. 123--134.

\bibitem{kuzu2013efficient}
M.~Kuzu, M.~Kantarcioglu, A.~Inan, E.~Bertino, E.~Durham, and B.~Malin,
  ``Efficient privacy-aware record integration,'' in \emph{Proceedings of the
  16th International Conference on Extending Database Technology}, 2013, pp.
  167--178.

\bibitem{cao2015hybrid}
J.~Cao, F.-Y. Rao, E.~Bertino, and M.~Kantarcioglu, ``A hybrid private record
  linkage scheme: Separating differentially private synopses from matching
  records,'' in \emph{2015 IEEE 31st International Conference on Data
  Engineering}.\hskip 1em plus 0.5em minus 0.4em\relax IEEE, 2015, pp.
  1011--1022.

\bibitem{he2017composing}
X.~He, A.~Machanavajjhala, C.~Flynn, and D.~Srivastava, ``Composing
  differential privacy and secure computation: A case study on scaling private
  record linkage,'' in \emph{Proceedings of the 2017 ACM SIGSAC Conference on
  Computer and Communications Security}, 2017, pp. 1389--1406.

\bibitem{quick2021generating}
\BIBentryALTinterwordspacing
H.~Quick, ``Generating poisson-distributed differentially private synthetic
  data,'' \emph{Journal of the Royal Statistical Society: Series A (Statistics
  in Society)}, vol. 184, no.~3, pp. 1093--1108, 2021. [Online]. Available:
  \url{https://rss.onlinelibrary.wiley.com/doi/abs/10.1111/rssa.12711}
\BIBentrySTDinterwordspacing

\bibitem{quick2021comparison}
H.~Quick, K.~Chen, and D.~DeLara, ``Comparison of poisson-gamma and laplace
  mechanisms for differential privacy,'' in \emph{2021 Workshop on Theory and
  Practice of Differential Privacy}, 2021.

\bibitem{quick2021improving}
H.~Quick, ``Improving the utility of poisson-distributed, differentially
  private synthetic data via prior predictive truncation with an application to
  cdc wonder,'' \emph{arXiv preprint arXiv:2103.03833}, 2021.

\bibitem{machanavajjhala2008privacy}
A.~Machanavajjhala, D.~Kifer, J.~Abowd, J.~Gehrke, and L.~Vilhuber, ``Privacy:
  Theory meets practice on the map,'' in \emph{2008 IEEE 24th international
  conference on data engineering}.\hskip 1em plus 0.5em minus 0.4em\relax IEEE,
  2008, pp. 277--286.

\end{thebibliography}
